\documentclass[lettersize,journal]{IEEEtran}
\usepackage[mathscr]{euscript}
\usepackage{color}
\usepackage{optidef}
\usepackage[algo2e,ruled,vlined]{algorithm2e}
\usepackage{float}
\usepackage{graphicx}
\usepackage{tikz,pgfplots}
\usepackage{upgreek}
\usepackage{multirow}
\usepackage{yfonts}
\usepackage{mathtools}
\usepackage[normalem]{ulem}
\usepackage{latexsym}
\usepackage{url}
\usepackage{amsmath,amssymb,amsbsy,amsfonts}
\usepackage{mathrsfs}
\usepackage{bbm}
\usepackage[export]{adjustbox}

\newcommand\restr[2]{{
  \left.\kern-\nulldelimiterspace % automatically resize the bar with \right
  {#1}\vphantom{\big|} \right|_{#2}}}

\DeclareMathOperator*{\argmin}{argmin} 
\newcommand{\R}{\mathbb{R}}

\newcommand{\data}{ \bs d }

\newcommand{\bs}[1]{\ensuremath{\boldsymbol{#1}}}

\usepackage{xspace}

\newcommand{\rr}{\ensuremath{\boldsymbol r}}

\newcommand{\sos}{c}

\newcommand{\source}{s}

\newcommand{\nn}[1]{\Phi_{#1}}
\newcommand{\wts}{\bs \xi}

\usepackage{cite}
\usepackage{amsmath,amssymb,amsfonts}
\usepackage{mathtools}
\usepackage{graphicx}
\usepackage{textcomp}
\def\BibTeX{{\rm B\kern-.05em{\sc i\kern-.025em b}\kern-.08em
    T\kern-.1667em\lower.7ex\hbox{E}\kern-.125emX}}
\usepackage[math]{cellspace}
\usepackage{graphicx}
\usepackage{algorithm}
\usepackage{algpseudocode}

\usepackage{float}
\usepackage{enumerate}
\usepackage{url}
\usepackage{multirow}

\usepackage{wrapfig,lipsum,booktabs}

\pgfplotsset{compat = 1.17}
\begin{document} 
\title{Learned Full Waveform Inversion Incorporating Task Information  for Ultrasound Computed Tomography}

\author{Luke Lozenski, \IEEEmembership{Student Member, IEEE}, Hanchen Wang, \IEEEmembership{Member, IEEE}, Fu Li \IEEEmembership{Student Member, IEEE},\\ 
Mark Anastasio \IEEEmembership{Senior Member, IEEE},
Brendt Wohlberg, \IEEEmembership{Fellow, IEEE},\\ Youzuo Lin, \IEEEmembership{Member, IEEE}, and Umberto Villa
%\thanks{Department of Electrical and Systems Engineering,
 %Washington University in St$.\,$Louis, St$.\,$Louis, MO 63130, USA}
\thanks{Luke Lozenski is with both the Department of Electrical and Systems Engineering, Washington University in St$.\,$Louis, St$.\,$Louis, MO 63130, USA and the Energy and Natural Resources Security Group, Los Alamos National Laboratory, Los Alamos, NM  87545, USA. Hanchen Wang and Youzuo Lin are with the Energy and Natural Resources Security Group, Los Alamos National Laboratory, Los Alamos, NM  87545, USA. Fu Li and Mark Anastasio are with the Department of Bioengineering, University of Illinois Urbana-Champaign, Urbana, IL, 61801 USA.
Brendt Wohlberg is with the Applied Mathematics and Plasma Physics Group, Los Alamos National Laboratory, Los Alamos, NM  87545, USA. Umberto Villa is with the Oden Institute for Computational Engineering and Sciences, University of Texas at Austin, Austin, TX 78712.}

\thanks{Further author information: (Send correspondence to Umberto Villa.)\\ E-mail: uvilla@oden.utexas.edu, Telephone  512-232-3453}}

%uvilla@wustl.edu, Telephone: +1 314 935 4777}}



\newcommand{\ylremark}[1]{\textcolor{blue}{[\textit{#1}]}}

\maketitle
\vspace{-.11cm}
\begin{abstract}

Ultrasound computed tomography (USCT) is an emerging imaging modality that holds great promise for breast imaging. Full-waveform inversion (FWI)-based image reconstruction methods incorporate accurate wave physics to produce high spatial resolution quantitative images of speed of sound or other acoustic properties of the breast tissues from USCT measurement data. However, the high computational cost of FWI reconstruction represents a significant burden for its widespread application in a clinical setting. The research reported here investigates the use of a convolutional neural network (CNN) to learn a mapping from USCT waveform data to speed of sound estimates. The CNN was trained using a supervised approach with a task-informed loss function aiming at preserving features of the image that are relevant to  the detection of lesions. A large set of anatomically and physiologically realistic numerical breast phantoms (NBPs) and corresponding simulated USCT measurements was employed during training. Once trained, the CNN can perform real-time FWI image reconstruction from USCT waveform data. The performance of the proposed method was assessed and compared against FWI using a hold-out sample of 41 NBPs and corresponding USCT data. Accuracy was measured using relative mean square error (RMSE), structural self-similarity index measure (SSIM), and lesion detection performance (DICE score). This numerical experiment demonstrates that a supervised learning model can achieve accuracy comparable to FWI in terms of RMSE and SSIM, and better performance in terms of task performance, while significantly reducing computational time.

\end{abstract} 

\begin{IEEEkeywords}
Ultrasound Computed Tomography, Convolutional Neural Networks, Data-Driven Image Reconstruction, Task Informed Image Reconstruction, Computer-simulation Study
\end{IEEEkeywords}

\section{Introduction}

Ultrasound computed tomography (USCT) is an emerging medical imaging technology that can provide high-resolution estimates of tissue acoustic properties by utilizing ultrasound and tomographic principles. Image formation in USCT is based on the interaction of acoustic wave signals with biological tissues. Quantitative reconstructions of a tissue's acoustic properties from USCT data can then be achieved via a variety of computational methods~\cite{Koulountzios21}, providing high-resolution images of breast tissue acoustic properties 
of significant diagnostic value for breast cancer \cite{duric2007detection, gemmeke07, hopp16}.

Full waveform inversion (FWI) is an image reconstruction method that estimates high-resolution maps of breast tissue acoustic properties from measurements of pressure distributions.  FWI models the propagation of an ultrasound signal in biological tissues by numerical solution of wave equation. By incorporating accurate wave-physics in the implementation of the imaging operator, FWI allows for superior accuracy and resolution compared to geometric reconstruction methods for USCT, such as bent-ray methods \cite{wang2015waveform, guasch20, huang14, pratt07, javaherian2021ray, hormati2010robust}. However, this comes at the cost of a significant computational burden compared to geometric reconstruction methods.  A single 3D reconstruction can take hours or days to compute and requires a high-performance, possibly GPU accelerated, computer\cite{lucka21, zhang12}.  This computational expense is a limiting factor for the widespread applications of FWI in a clinical setting where fast reconstruction methods are highly desired. Furthermore, the need for a powerful computer increases the cost of USCT and prevents its adoption in developing areas.

This work proposes a learned FWI method utilizing convolutional neural networks (CNNs) for accelerated reconstructions. Neural networks have demonstrated the ability to construct inverse mappings of nonlinear imaging operators~\cite{wang20, reader20}, including some promising methods for USCT reconstruction~\cite{fan21, donaldson21, prasad22, liu21,jeong2023deep}. A key contribution of this work is to acknowledge and leverage the fact that USCT is often used for specified diagnostic tasks, such as tumor detection and localization. To this aim, a novel loss function is proposed that includes task-specific information using a model of tumor segmentation provided by a U-Net numerical observer~\cite{ronneberger2015u}. Specifically, the proposed loss function consists of a weighted sum of the commonly-used mean square error loss in the image domain and a novel task-informed objective based on features extracted by the U-Net observer. A second contribution of this work is the use of source encoding, a common technique used in FWI to reduce computational cost, within a learned reconstruction method. By exploiting redundancies in data, source encoding can reduce complexity of the CNN architecture and accelerate training of the learned FWI method. It is also worth noting that the proposed method is developed with the use of anatomically realistic training and testing sets that are relevant to breast imaging~\cite{badano2018evaluation, PhantomGen}.

Four simulation studies are performed to demonstrate the feasibility of the proposed method. The first study assesses the role of source encoding as a means of data reduction. Source encoding involves the  application of linear combinations of sources and corresponding measurement data to reduce the overall size and complexity of the learned FWI network.  This study compares different source encoding methods and assesses the resulting accuracy of the learned reconstruction methods after a fixed number of training epochs. The second study assesses the role of incorporating task-specific information into a learned reconstruction method. Here, the specific task chosen is tumor detection and localization. Task information is incorporated into training with a task-informed objective function, which includes a term based on features extracted from a U-Net observer. In this study, multiple learned reconstruction methods are trained using a sequence of loss functions that pose increased emphasis on the task-informed loss. The third study assessed the robustness of the learned reconstruction with respect to noise. In this study, the learned reconstruction methods constructed in the second study were reassessed using measurements with a higher level of added noise than they were trained on.  The fourth study assessed the generalizability of the learned reconstruction method. In this study, the training and testing sets were separated into distinct distributions and the accuracy of the trained CNN was assessed for reconstruction of underrepresented populations in the training set.  

The remainder of this paper is structured as follows. In Section \ref{sec:background}, the USCT imaging operator in its continuous and discretized form as well as the full waveform inversion (FWI) method are reviewed. This section also provides a brief discussion on task-based assessment of image quality  and its role in USCT. In Section \ref{sec:method}, a solution method utilizing a specific CNN architecture, InversionNet \cite{inversionNET}, is presented with particular emphasis on the implementation of source encoding and the proposed task-informed loss function. In Section \ref{sec:numerical_studies}, the design of four numerical studies is presented: the first study compares multiple methods of source encoding for use with InversionNet; the second study assesses the use of a  task-informed loss function with varying weights on task information; the third study assesses the proposed methods robustness with measurement noise; the fourth study analyzed the ability of the proposed method to reconstruct new unseen objects from an underrepresented population in the training set. In Section \ref{sec:results}, the results of the numerical studies are presented. Section \ref{sec:conclusion} presents the conclusions drawn from these results and discusses future extensions.

\section{Background} \label{sec:background}

\subsection{Imaging Operator} USCT data is formed by measuring acoustic signals resulting from a series of ultrasonic excitation pulses emitted from multiple tomographic views surrounding the object to be imaged. These ultrasound waves generated from these excitation pulses then interact with the object according to the tissues' acoustic properties. The reflected and scattered ultrasound waves then propagate outside of the object where they are recorded by a set of receiving transducers surrounding the object, often with water as a coupling medium between the object and the measurement surface. 

In this work, the USCT imaging operator is modeled by solving the acoustic wave equation.  This work assumes a nonlossy medium with homogeneous density \cite{jensen91} and a spatially varying speed of sound $\sos = \sos(\rr)$, where $\rr\in \R^d$ ($d = $ 2 or 3) is a point in the spatial domain. For pressure measurements collected on an aperture $\mathcal{S}$ surrounding the object, the relationship between the measurements and the speed of sound $\sos$ can be expressed as a continuous to continuous (C-C) imaging operator defined as
\begin{equation}\label{eq:cc}
\mathcal{H}^\sos \source_i \coloneqq p_i(\rr,t ) \quad  (\rr,t) \in \mathcal{S} \times [0,T],
\end{equation} for $i=1,\ldots, I$ where $T$ denotes the acquisition time for a single shot, $I$ denotes the total number of emitting transducers, $s_i = s_i(\rr, t) $ and $p_i=p_i(\rr,t )$ denote the excitation pulse and the acoustic pressure field generated when the $i$-th emitter is fired. Above, the notation $\mathcal{H}^\sos$ is used to underline the dependence of the imaging operator on the medium speed of sound $\sos$. Under the assumption of a non-lossy propagation medium with homogeneous density, the acoustic pressure field $p_i$ satisfies the wave equation
\begin{equation}\label{eqn:waveq}
\begin{array}{cc}
    \frac{1}{c(\rr)^2}\frac{\partial^2}{\partial t^2} p_i(\rr,t) - \Delta p_i(\rr,t) = s_i(\rr,t)  & (\rr,t) \in \R^d \times [0,T] \\
     p_i(\rr,0) = 0& \rr\in \R^d \\ 
     \frac{\partial}{\partial t}p_i(\rr,0) = 0& \rr\in \R^d . 
\end{array}
\end{equation}

Assuming that $J$ idealized point-like transducers are distributed along the measurement aperture at locations $\rr_j \in \mathcal{S}$ ($j=1, \ldots J$), the sampling operator $\mathcal{M}$ mapping the pressure $p(\rr,t )$ to the vector $\boldsymbol{g} \in \mathbb{R}^{K J}$ is defined as 
\begin{equation}
 [\mathcal{M} p]_{k+(j-1)K} \coloneqq [\boldsymbol{g}]_{k+(j-1)K} = p(\rr_j, k\Delta T),
\end{equation}
where  $k = 1,\hdots, K$; $\Delta T = T/K$ is the sampling interval; and $K$ is the number of pressure samples measured over the acquisition interval $[0,T]$.

Using the continuous-to-discrete imaging operator $\mathcal{M}\mathcal{H}^c$, the USCT data acquisition process is modeled as 
\begin{equation}
\boldsymbol{d}_i = \mathcal{M}\mathcal{H}^c s_i + \boldsymbol{n}_i \quad i = 1, \ldots, I,
\end{equation}
where %$s_i := s_i(\rr, t)$ is the $i^{\rm th}$ excitation pulse, $I$ is the number of sources, and 
$\boldsymbol{n}_i\in \mathbb{R}^{K \times J}$ is additive noise \cite{Li23}.

Finally, with the introduction of a Cartesian grid consisting of $Q$ pixels, the imaging operator can be approximated with an analogous discrete-to-discrete (D-D) imaging operator. Denoting the center of the $q^{\rm th}$ pixel with $\boldsymbol{r}_q$, the finite-dimensional vectors $\boldsymbol{c} \in \mathbb{R}^Q$ and $\boldsymbol{s}_i \in \mathbb{R}^{K J}$ are defined as 
\begin{equation}\label{eqn:discretization}
    \begin{array}{c}
   [\boldsymbol{c}]_q = c(\rr_q), \; [\boldsymbol{s}_i]_{k+(q-1)K} = s_i(\rr_q, k\Delta T)  \\
     \quad q=Q, \ldots, N; \, k=1,\ldots, K. 
\end{array}
\end{equation} 

With the above notation, the D-D USCT model is given by
\begin{equation}\label{eqn:imaging}
\boldsymbol{d}_i = \boldsymbol{M}\boldsymbol{H}^{\boldsymbol{c}} \boldsymbol{s}_i + \boldsymbol{n}_i \quad i = 1, \ldots, I,
\end{equation}
where $\boldsymbol{M}: \mathbb{R}^{K Q} \mapsto \mathbb{R}^{K J}$ is the discrete counterpart of the sampling operator $\mathcal{M}$ defined via nearest neighbor interpolation of transducer coordinates to the pixel centers of the Cartesian grid, $\boldsymbol{H}^{\boldsymbol{c}}:  \mathbb{R}^{KQ} \mapsto \mathbb{R}^{K Q}$ stems from finite difference approximation of the C-C imaging operator $\mathcal{H}^\sos$, and $\boldsymbol{n}_i$ ($i = 1,\ldots,I$) represents measurement noise.

\subsection{Full Waveform Inversion}

Full waveform inversion (FWI) ~\cite{virieux2009overview} is a method for high-resolution reconstruction of the speed of sound map $\sos$ given pressure traces $\data$ popularly  used in the geophysics community. FWI utilizes the D-D imaging model in  Eq. \eqref{eqn:imaging} and seeks a speed of sound estimate $\hat{\boldsymbol{c} }$ such that
\begin{equation}
\hat{\boldsymbol{c} } := \argmin_{\boldsymbol{\sos} \in \R^Q} \frac{1}{2}\sum_{i=1}^I \left\| \data_i - \boldsymbol{M}\boldsymbol{H}^{\boldsymbol{c}} \boldsymbol{s}_i  \right\|^2.
\label{eq:fwi_det}
\end{equation}
This discrete optimization problem can be solved using a gradient-based method to update estimates of $\boldsymbol{\sos}$. 
However, each evaluation of the objective function and its gradient require the solution of $I$ forward and adjoint wave equations. \emph{Source encoding} is a technique that, by leveraging the linearity of the imaging operator $\mathcal{H}^c$ with respect to the excitation pulse $s$, allows for a drastic reduction in the computational cost \cite{krebs09, wang2015waveform, lucka21}. In the source encoding method, the deterministic minimization problem in Eq. \eqref{eq:fwi_det} is reformulated as the stochastic optimization problem 
%\begin{equation}
%\hat{\boldsymbol{c} } := %\argmin_{\boldsymbol{\sos} \in \R^Q} %\frac{1}{2}  \mathbb{E}_{w_{i} \sim \Gamma} %\left\| \sum_{i=1}^I w_i(\data_i -  %\boldsymbol{M}%\boldsymbol{H}^{\boldsymbol{c}} %\boldsymbol{s}_i ) \right\|^2,
%\end{equation} where each $\{w_i\}_{i=1}^I$ is a random variable drawn from a distribution $\Gamma$, which has zero mean and unit variance.
\begin{equation}
\hat{\boldsymbol{c} } := \argmin_{\boldsymbol{\sos} \in \R^Q} \frac{1}{2}  \mathbb{E}_{\boldsymbol{w} \sim \Gamma} \left\| \data_{\boldsymbol{w}} -  \boldsymbol{M}\boldsymbol{H}^{\boldsymbol{c}} \boldsymbol{s}_{\boldsymbol{w}} \right\|^2,
\label{eq:wise}
\end{equation}
where $\boldsymbol{w} \in \mathbb{R}^I$ is the stochastic encoding vector sampled according to the distribution $\Gamma$ with zero mean and identity covariance matrix, and $\data_{\boldsymbol{w}} = \sum_{i=1}^I [\boldsymbol{w}]_i \data_i$, $\boldsymbol{s}_{\boldsymbol{w}} = \sum_{i=1}^I [\boldsymbol{w}]_i \boldsymbol{s}_i$ denote the superimposed (encoded) measurement data and excitation source, respectively. Previous studies have used a Rademacher or a normal distribution to sample the independent identically distributed components $[\boldsymbol{w}]_i$ of the encoding vector. The stochastic optimization problem in Eq. \eqref{eq:wise} is then often solved using stochastic gradient descent, thus requiring the solution of only one forward and one backward wave equation for iteration. However, even when acceleration techniques, such as Nesterov \cite{nesterov1983method} or momentum \cite{polyak1964some}, or modern stochastic optimization methods, such as ADAM \cite{KingmaBa2014}, are applied,  convergence is still slow and reconstructions for individual images may take several minutes or hours.

\subsection{Task-Based Assessment of Image Quality}

In biomedical imaging, the reconstructed image is often utilized to inform a specific task, with the image itself being of secondary interest. Such tasks can include but are not limited to, classification, segmentation, registration, detection, and decision planning. However, in many cases, metrics of task performance may not be directly correlated to physical metrics of image quality~\cite{barrett93, christianson15}, such as the mean square error or the structural similarity index. Furthermore, designing an image reconstruction strategy focusing purely on image quality may have the unintended consequence of reducing task performance \cite{li21}. For improved task performance, task-based information must be included in the design of an image reconstruction strategy \cite{zhang21, adler22}. These tasks are often automated using a numerical observer, which are mathematical models that identify the task-relevant features in an image and estimate the resulting task outcome\cite{he2013model}. Examples of numerical observers include Luenberger observers and Kalman filters for control tasks\cite{luenberger1971introduction}, library and model-based dose estimation for treatment planning tasks \cite{moore2019automated}, and Hotelling observers (or channelized versions) and machine learning based observers for signal detection tasks \cite{li2023estimating}. With a numerical observer, task based-assessment of large sets of images can be done quickly and used to design a reconstruction method for optimal task performance.

\section{Method}\label{sec:method}

This section presents the main contribution of this work: the development of a task-aware learned USCT reconstruction method utilizing CNNs.  Once fully trained, this CNN acts as an inverse mapping from the set of pressure traces to the corresponding speed of sound map. The specific CNN architecture used here is InversionNet \cite{inversionNET}, which was originally developed for FWI of seismic waveform data in geophysics. Both seismic and USCT imaging problems are focused on reconstructing speed of sound based on acoustic wave models but have a few key differences.  First, measurements in seismic imaging are sparse and expensive to acquire while USCT measurements can be quite dense. This work explores utilizing source encoding for reducing data complexity in USCT data and accelerating training. Second, seismic and USCT have very different image priors. A learned USCT reconstruction method then requires application-relevant training and testing sets. This work utilizes medically realistic, stochastically generated acoustic breast phantoms to construct the training and testing sets \cite{PhantomGen}. Third, the end goal for seismic imaging in geophysics is often to understand geological structures based on acoustic properties, whereas USCT imaging is often focused on a medically specific task with structural information being a secondary concern.  This work then explores utilizing a task-based objective function to train a learned reconstruction method tailored for tumor/lesion detection and localization.

\subsection{Learned FWI via InversionNet}

InversionNet utilizes an end-to-end trained encoder-decoder structure. In this scheme, pressure traces are encoded to a high-dimensional latent space and then decoded to the space of images. Specifically, the input to InversionNet is a 3-D tensor $\boldsymbol{D} \in \R^{I \times K \times J}$ where $[\boldsymbol{D}]_{ijk} = [d_i]_{k + (j-1)K}$  corresponds to the measurement data from the $i$-th source, $j$-th receiver, and $k$-th time sample and its output is a 2-D tensor $\boldsymbol{C} \in \mathbb{R}^{Q_x \times Q_y}$ (with $Q = Q_x \times  Q_y$) corresponding to pixel values of speed of sound estimates over the field of view.

The parameters $\wts \in \R^W$ of the InversionNet $\nn \wts $ are then  trained by minimizing the loss function 
\begin{equation}\label{eqn:supervised}
\min_{\wts \in \R^p} \frac{1}{2N} \sum_{n=1}^N\| \nn \wts (\boldsymbol{D}^n) - \boldsymbol{C}^n\|^2,
\end{equation} where $\{(\boldsymbol{C}^n,\boldsymbol{D}^n)\}_{j=1}^N$ are data pairs consisting of the speed of sound maps and corresponding USCT measurements. Here the $\ell^2$ norm is implemented in the training loss; however, other choices of loss are possible \cite{inversionNET}.

\subsection{Source Encoding}

USCT data often contains several redundancies %in terms of spatial information 
and a large memory footprint for measurements collected at multiple receivers over a long sampling period when multiple sources are sequentially excited one at a time. Furthermore, InversionNet was originally designed for seismic reconstruction in geophysics in which measurements are collected with sparse spatial coverage. Nevertheless, InversionNet demonstrates sufficient performance on these sparse measurements. This observation, together with the large success in applying source encoding methods to accelerate FWI reconstruction of USCT data, is the key motivation to explore methods to reduce the dimensionality of the measurement data provided as input to InversionNet.

This work proposes utilizing a fixed source encoding approach to reduce the dimensionality of the data. Exploiting redundancies in the data, not only allows for to reduce memory footprint of InversionNet, but it also has the potential of improving the performance of the learned method~\cite{chen23}.

As described in Section \ref{sec:background}, source encoding, originally proposed for accelerated FWI in geophysics problems \cite{krebs09}, exploits the linearity of the imaging operator with respect to the source term. Given an encoding vector $\boldsymbol{w} \in \mathbb{R}^I$, the superimposition of multiple sources in the D-D USCT model in Eq. \eqref{eqn:imaging} gives 
\begin{equation}
\boldsymbol{d}_{\boldsymbol{w}} = \boldsymbol{M}\boldsymbol{H}^{\boldsymbol{c}} \boldsymbol{s}_{\boldsymbol{w}} + \boldsymbol{n}_{\boldsymbol{w}},
\end{equation}
where $\boldsymbol{d}_{\boldsymbol{w}} = \sum_{i=1}^I [\boldsymbol{w}]_i \boldsymbol{d}_i$, $\boldsymbol{s}_{\boldsymbol{w}} = \sum_{i=1}^I [\boldsymbol{w}]_i \boldsymbol{s}_i$, and $\boldsymbol{n}_{\boldsymbol{w}} = \sum_{i=1}^I [\boldsymbol{w}]_i \boldsymbol{n}_i$ denote the superimposed (encoded) measurement data, excitation pulse, and measurement noise, respectively.
%\begin{equation}
%\sum_{i=1}^I w_i \boldsymbol{d}_i \approx  \boldsymbol{M}\boldsymbol{H}^{ \boldsymbol c} \sum_{i=1}^I w_i s_i,
%\end{equation} where $\{w_i\}_{i=1}^I \subset \R$.
By use of $L < I$ independent source encoding vectors $\boldsymbol{w}_l$, the input $\boldsymbol{D} \in \R^{I \times K \times J}$ can be reduced to a smaller tensor $\boldsymbol{D}_{\boldsymbol{W}} \coloneqq {\boldsymbol{W}}{\boldsymbol{D}} \in \R^{L \times K \times J}$, where $\boldsymbol{W} \in \R^{L \times I}$ is the encoding matrix with entries $[\boldsymbol{W}]_{li} = [\boldsymbol{w}_l]_i$. Above, $\boldsymbol{D}_{\boldsymbol{W}} = {\boldsymbol{W}}{\boldsymbol{D}}$ denotes the matrix-tensor multiplication defined by saturation of the last index of the matrix  $\boldsymbol{W}$ with the first index of the 3D tensor ${\boldsymbol{D}}$. That is, the entries of $\boldsymbol{D}_{\boldsymbol{W}}$ are given by
\begin{equation}
[\boldsymbol{D}_{\boldsymbol{W}}]_{lkj} = \sum_{i=1}^I [\boldsymbol{W}]_{li} [\boldsymbol{D}]_{ikj}, 
\end{equation}
for $l=1,\ldots,L$; $k=1,\ldots,K;$ $j=1,\ldots,J$.

%Reducing the number of input channels reduces computational complexity in training InversionNet by a factor of $\frac{I}{L}$.
For a given (fixed) encoding matrix $\boldsymbol{W}$, InversionNet is then trained by solving the minimization problem 
$$\min_{\wts \in \R^p} \frac{1}{2N} \sum_{n=1}^N\| \nn \wts ({\boldsymbol{W}}\boldsymbol{D}) -\boldsymbol  C^n\|^2.$$  
%Learned data-augmentation methods have demonstrated modest improvements in the performance of some deep learning methods \cite{cubuk19}.
An instance of InversionNet with a learned sourced encoder is also considered and was trained by solving the minimization problem
$$\min_{\wts \in \R^p, \boldsymbol{W} \in \R^{L\times I}} \frac{1}{2N} \sum_{n=1}^N\| \nn \wts ({\boldsymbol{W}}\boldsymbol{D}) - \boldsymbol C^n\|^2.$$

\subsection{Task-Informed Training}

Medical imaging is typically performed for some specific diagnostic task. A learned reconstruction method should therefore be assessed and optimized with respect to this task.  In particular, it is desirable to develop learned reconstruction methods that can preserve task-relevant information in the reconstructed images. This can be achieved by utilizing a task-informed loss function during training. To this aim, the following supervised training problem is considered 
\begin{equation}\label{eqn:task_informed_obj}
\begin{split}
\mathop{\min}_{\wts \in \R^p}  \frac{1}{2N} \sum_{n=1}^N & \left\{\| \nn \wts (\boldsymbol{W}\boldsymbol{D}^n) - \boldsymbol{C}^n\|^2 + \right.\\
+ & \left. \gamma \| T(\nn \wts (\boldsymbol{W}\boldsymbol{D}^n)) - \boldsymbol{T}(\boldsymbol C^n) \|^2\right\},
\end{split}
\end{equation} where the first term is the MSE loss and the second term is the task-informed loss. Above, $\gamma \geq 0$ is a weighting factor balancing the trade-off between the two losses, and $\boldsymbol{T}$ is a differentiable (possibly) non-linear map that extracts the task-relevant features of the image. 
%Here $T$ acts as a differentiable model of the specified task. This means that if $T(\nn \wts (\data^n)) = T(\boldsymbol\sos^n)$, then $\nn \wts (\data^n)$ and  $\sos^n$ have identical task performance.
The map $T$ can then take a variety of forms depending on the task and should be designed based on the numerical observer used. For example $T$ could be a linear mapping, such as the template of a Hotelling observer \cite{barrett2013foundations} or a projector onto the linear subspace spanned by channels of a channelized Hotelling observer \cite{yao1992predicting, gifford2000channelized}, or a more general nonlinear mapping between an image and task-relevant features, possibly learned by a CNN-based approximation of the ideal observer \cite{li2023estimating}.

Setting the task weighting factor $\gamma$ to $0$
 recovers the original supervised training problem and creates an image reconstruction strategy only focused on physical metrics of image quality.  Increasing $\gamma$ places a greater emphasis on task performance. Setting $\gamma = \infty$ (which means the MSE component is discarded) creates an image reconstruction strategy that is only driven by the task. 

 However, even though image quality and task performance are not directly linked~\cite{zhang21, li21}, there is still an indirect relationship. Therefore an image reconstruction strategy that has the proper weighting between MSE and task information, has the potential to outperform, both in terms of image quality and task performance, the reconstruction strategies that are only informed by image quality or task information. This means that for a properly chosen task there is a selected value of $\gamma \in(0,\infty)$ that maximizes both image quality and task performance. 
 It should be noted that $T$ may have a null-space or the nonlinearity of $T$ may present a harder optimization problem to solve with large values of $\gamma$. In practice this issue can be alleviated by training with successively increasing values of $\gamma$. This training process then allows the MSE component to first drive the solution towards a good local minima in terms of physical metrics of image quality and then progressively incorporate more task information.

\section{Numerical Studies}\label{sec:numerical_studies}

\subsection{Construction of the Training and Testing Sets}

\begin{figure*}[tbh]
    \centering
    \resizebox{\textwidth}{!}{ \includegraphics[trim = {2cm, 3cm, 7.9cm 2cm}, clip]{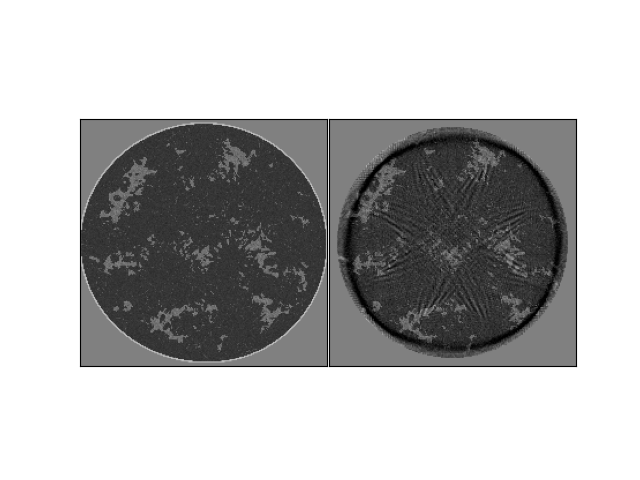}\includegraphics[trim = {2cm, 3cm, 7.9cm 2cm}, clip]{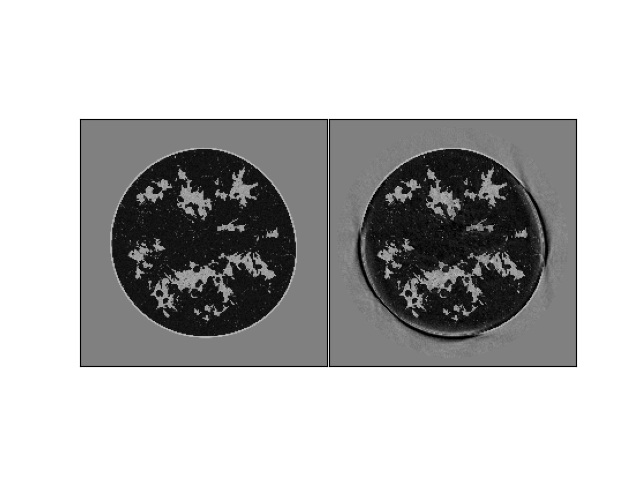}\includegraphics[trim = {2cm, 3cm, 7.9cm 2cm}, clip]{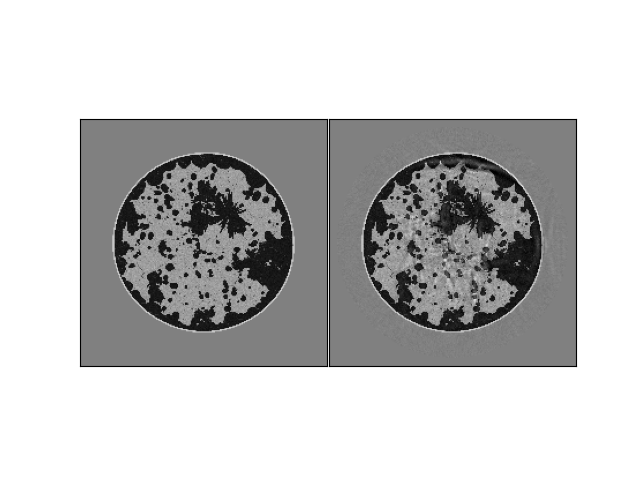}\includegraphics[trim = {2cm, 3cm, 7.9cm 2cm}, clip]{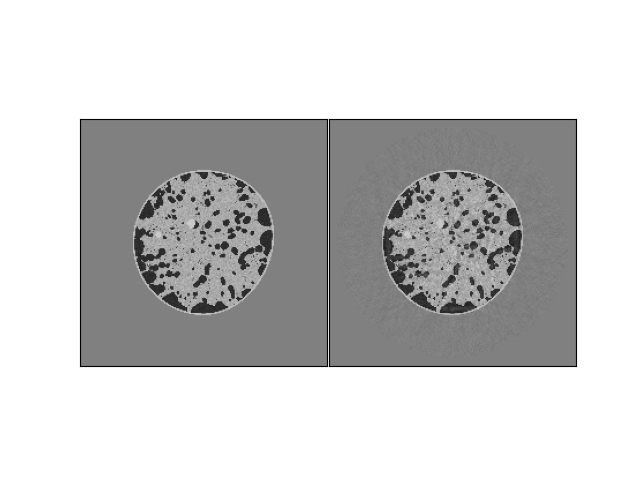}}
    \caption{Four examples of the anatomically realistic numerical breast phantoms (NBPs), one from each of BI-RADS breast density types, used to train and evaluate the proposed learned FWI method. NBPs present clinically relevant variability in size, tissue composition and structures, and speed-of-sound maps. From left to right: Type A (almost all fatty), Type B (scattered fibroglandular density), Type C (heterogeneous density), and Type D (extremely dense). }
    \label{fig:examples_NBPs}
\end{figure*}

The objects used in this study were anatomically realistic numerical breast phantoms (NBPs) to which spatially varying speeds of sound were stochastically assigned within feasible ranges. These NBPs were developed and constructed by Li et al \cite{PhantomGen} using tools adapted from the Virtual Imaging Clinical Trial for Regulatory Evaluation (VICTRE) project at the US Food and Drugs Administration \cite{badano2018evaluation} for use in USCT virtual imaging studies. Examples of these NBPs are available from \cite{LiDataverse2021}. In particular, the  generated NBPs are stratified based on the four different levels of breast density (percentage of fibroglandular tissue) defined according to the American College of Radiology's (ACR) Breast Imaging Reporting and Data System (BI-RADS) \cite{american2013acr}: A] almost all fatty breasts, B] breasts with scattered fibroglandular density, C] breasts with heterogeneous density, and D] extremely dense breasts. The breast size and percentage of fibroglandular tissue of each NBPs are randomly chosen based on the within the physiological range for each breast density type. Then anatomically realistic breast tissue structures are stochastically generated by the use of the VICTRE tools and the speed of sound maps corresponding to physiological variations in breast tissues are stochastically generated \cite{PhantomGen}. Four NBPs, one from each of the BI-RADs categories, are shown in Fig. \ref{fig:examples_NBPs}. The training set consisted of 1,353 NBPs while the testing test consisted of 41 NBPs.

\subsection{Definition of the Virtual Imaging System} \label{sec:d2d}
\begin{table*}
\caption{Virtual imaging system and discretization parameters}% \umbe{If the number of points per wavelength state here is correct, the output of the wavesolver may be severely inaccurate. Ideally, finite difference need 10 points per wavelength.}}
    \centering
    \begin{tabular}{ll | ll }
    \hline
    \multicolumn{2}{c|}{Ultrasound system} & \multicolumn{2}{c}{Discretization} \\
        Number of receivers $J$ & 256 & Grid size $N_x$ & 360 \\ 
        Number of transmitters $I$ & 64 & Grid intervals $\delta x$ & 0.6 mm \\
        Transducer radius $R$ & 110.4 mm & Points per wavelength  $\frac{\sos_{min}}{f_0 \delta x}$ & 5.0\\ 
        Pulse frequency $f_0$ & $0.5$ MHz & Number of time steps $K$ & 640 \\  
        %Time of peak amplitude & $3.2 \mu$s \\
        %Signal width  $\sigma$ & 12 $\mu$s \\ 
        %Number of time steps $N_t$ & 400 \\ 
        %Time steps $\delta t$ & 0.2 $\mu$s\\ 
        Sampling frequency & $5$ MHz & Time steps $\delta t$ & 0.2 $\mu$s\\
        Acquisition time (per source) & 128 $\mu s$ & CFL Number  $\frac{\sos_{max} \delta t}{\delta x}$ & 0.53 \\
        \hline
        %Number of phantoms & 1,394 \\ 
        %Size of Training Set & 1,353\\ 
        %Size of Testing Set & 41 \\ 
    \end{tabular}
    \label{tab:ultrasound_system}
\end{table*}

%\begin{figure} \centering \includegraphics[trim = {5cm, 4.5cm 5cm 1cm}, clip, width = 0.45 \textwidth]{Figures/usct_setup.png} \caption{Illustration of the virtual 2D imaging system. A total of 256 transducers are arranged in circular array.  Transducers marked in red act as both receivers and emitters, transducers marked in blue act only as receivers. The field of view (FOV) where the speed of sound map is estimated is outlined in green.} \label{fig:imaging_system} \end{figure}

%Figure \ref{fig:imaging_system} illustrates the 2D virtual imaging system used in the numerical studies. 

The measurements geometry consisted of a circular transducer array\cite{wang2015waveform, stotzka2002medical} of radius $R$ along which 256 transducers (shown in blue and red) were equispaced and acted as receivers. Every fourth transducer (shown in red), 64 in total, also acted as a transmitter and would emit an excitation pulse in sequence. The $i$-th excitation pulse was of the form
$$\begin{array}{c}
     s_i(\rr,t) = \delta(\rr - \rr_i)\exp\left( -\frac{(t-t_0)^2}{2\sigma^2} \right) \sin(2\pi f_0 t)  \\
     i = 1,\hdots, I,
\end{array}$$
where $\rr_i$ is the location of the $i$-th emitter, $f_0$ is the central frequency, $t_0=3.2\mu$s is the time shift, and $\sigma=2\mu$s controls the signal width. Measurements are collected by firing one transmitter at a time and recording data at every receiver. This is repeated for each transmitter and results in multi-channel measurements.

To numerically simulate the pressure field generated by each transmitter, the wave equation in Eq. \eqref{eqn:waveq} was solved using a finite difference scheme (4th order in space and 2nd order in time). A spatial grid of size $N_x\times N_x$ ($Q = N_x^2$ in Eq. \eqref{eqn:discretization}), with $N_x=360$, and a temporal grid with $K = 400$ samples were employed for the discretization.  Absorbing boundary conditions were implemented to prevent wave reflections at the boundaries of the computational domain \cite{clayton1977absorbing}.  Electronic noise was modeled as additive white Gaussian noise  with a standard derivation of $4.3 \cdot 10^{-6},$ corresponding to an SNR of 30 dB. The imaging system parameters are summarized in Table \ref{tab:ultrasound_system}.

\subsection{Study Design}

\subsubsection{Study 1: Source encoding} The first numerical study addressed the role of source encoding in training InversionNet. This study investigates the use of source encoding as a method of data reduction to accelerate training. In this study, four different approaches for source encoding are considered. Four instances of InversionNet were compared: 1) The \emph{Reference} instance (117,762,947 trainable parameters) was trained with no source encoding and utilizing all 64 measurement channels; 2) The \emph{Subsample} instance (117,686,147 trainable parameters) was trained utilizing subsampling down to 16 measurement channels, i.e. only keeping every fourth measurement channel; 3) The \emph{Random} instance (117,686,147  trainable parameters) was trained with fixed randomly chosen source encoding, where the weights are drawn from a standard normal distribution, from  64 channels to 16 channels; 4) The \emph{Learned} instance (117,687,171 trainable parameters) was trained with by while jointly learning the weights of a source encoding from  64 channels to 16 channels. 

\subsubsection{Study 2: Task-informed loss} The second numerical study explored the impact of the task-informed loss function in training InversionNet, where the chosen task was tumor detection and localization. A numerical observer for this task was implemented by use of a U-net, a specific neural network architecture developed for the segmentation of medical images  \cite{ronneberger2015u}. The U-net is a multilevel architecture consisting of a contractive path, which applies a sequence of convolutional and downsampling layers to half the image size at each level, and an expansive path, which restores the output to its original size by use of transpose convolutional and upsampling layers. Skip connections are used to concatenate feature maps computed in the contractive path to the inputs of the corresponding layer in the expansive path. Specifically, the U-net-based numerical observer was  trained using the speed of sound maps as input and corresponding binary segmentation masks of tumor regions as outputs. More details can be found in Appendix \ref{sec:u-net-training}.
The task-relevant feature maps $T$ were then constructed by extracting and concatenating all feature maps computed as an output of the convolutional layers at each level in the contractive path of the U-Net.

Six versions of InversionNet were trained in succession, i.e., each network was initialized using the final trained state of the previous version. These instances of InversionNet were trained with varying task weights in the loss function in Eq. \eqref{eqn:task_informed_obj}. The selected task weights were $\gamma =0,10^{-3},10^{-2},10^{-1},1,10,\infty;$ where $\gamma =0$ corresponds to the original MSE loss function  and $\gamma = \infty$ corresponds to a task-only informed loss. The InversionNet reconstructions were then compared to a traditional FWI method on the testing set utilizing a source encoding stochastic descent method \cite{wang2015waveform}. 

\subsubsection{Study 3: Robustness w.r.t. noise}
The third numerical study explored the robustness of the learned reconstruction method to an increased amount of noise. In this study, the instances of InversionNet trained in the second study were used to reconstruct USCT data corrupted by \emph{i.i.d.} additive noise ten times larger than that used while training the network.  These reconstructions from measurements with increased noise were then compared to the results of the second numerical study for the same objects in the testing set. 

\subsubsection{Study 4: Generalizability w.r.t. underrepresented groups}
The fourth numerical study explored the generalizability of the proposed method when a group of objects is underrepresented in the training set. In particular, the anatomically realistic NBPs used to train and assess our method can be divided in four groups based on the corresponding BI-RADS breast density type. As shown in Fig. \ref{fig:examples_NBPs}, NBPs from different groups (density type) exhibit large differences in size and tissue composition. For this study, a training set with 1,377 NBPs (approximately the same number of training examples as that used in the previous study) was employed. However, the prevalence of type A (fatty breast) in the training set was only 6\%  (81 NBPs). An instance of InversionNet was then trained on this unbalanced training set as described in Study 2 and assessed using a testing set consisting of 81 NBPs from the underrepresented group for reconstructing underrepresented groups (BI-RADS breast density types) in the testing set.

\subsection{Image Quality Assessment Criteria}
Image quality was quantified in terms of ensemble average relative mean square error (RMSE) and ensemble average structural self-similarity index measure (SSIM) \cite{WangBoviketall04}. Task accuracy was quantified using by a numerical observer. In particular, the area under the curve (AUC) of the receiver operating characteristic (ROC) and a Dice coefficient for correct tumor detection and localization (\emph{tumor-wise} Dice coefficient) were employed. The ROC curve for each reconstruction method was constructed by plotting the tumor-wise true positive rate on the y-axis and the pixel-wise false positive rate on the x-axis for a range of thresholds. The AUC was then computed and used as a figure of merit to assess the proposed learned reconstruction method. 

The ROC curves were also used to select a threshold for tumor detection in the computation of the \emph{tumor-wise} Dice coefficient. In particular, the selected threshold parameter corresponds to the upper left corner of the ROC.
The \emph{tumor-wise} Dice coefficient was then computed as 
\begin{equation}\label{eqn:dice}
    \operatorname{Dice} = \frac{2 N_{\textnormal{True Detections}}}
{N_{\textnormal{Detections}} + N_{\textnormal{True Tumors}}},
\end{equation} where $N_{\textnormal{Detections}}$ is the number of detected tumors in the reconstructed image,  $N_{\textnormal{True Tumors}}$ is the number of tumors presented in the true image, and $N_{\textnormal{True Detections}}$ is the number of tumors that are correctly detected and localized by the numerical observer applied to the reconstructed image.

\section{Results and Discussion}\label{sec:results}

Simulation of USCT measurements and image reconstruction using the FWI with source encoding method were implemented using Devito \cite{kukreja2016devito}, a Python package for solving partial differential equation using optimized finite difference stencils. Each set of measurements was perturbed by additive white Gaussian noise with standard deviation $4.3\times10^{-6}$  (SNR = 30 dB) for the first, second, and fourth experiments while the third experiment utilized measurements with a ten times larger amount of noise (SNR = 20 dB).  The FWI reconstructions of the 41 test images used in the first three studies took approximately 6 hours on a workstation with two Intel Xeon Gold 5218 Processors (16 cores, 32 threads, 2.3 GHz, 22 MB cache each), 384 GB of DDR4 2933Mhz memory, and one NVidia Titan RTX 24GB graphic processing unit (GPU).

InversionNet was implemented in PyTorch, an open-source machine learning framework \cite{NEURIPS2019_9015}, and trained for 1,000 epochs with a batch size of 50 using Adam optimizer \cite{KingmaBa2014}. At each training iteration, noise with a fixed SNR of 30 dB was added online as a form of data augmentation, which is known to improve training \cite{shorten19}. Training each instance of InversionNet took approximately 10 hours on an HPC node with 512 GB of memory and 4 NVidia Volta V100 graphic processing units (GPUs). Evaluation of an instance of InversionNet on the testing set (41 examples) took approximately 1 second.

\subsection{Study 1: Source Encoding}

\begin{figure*}[tbh]
    \includegraphics[width = \textwidth, trim = {0cm 2cm 0cm 0cm},clip]{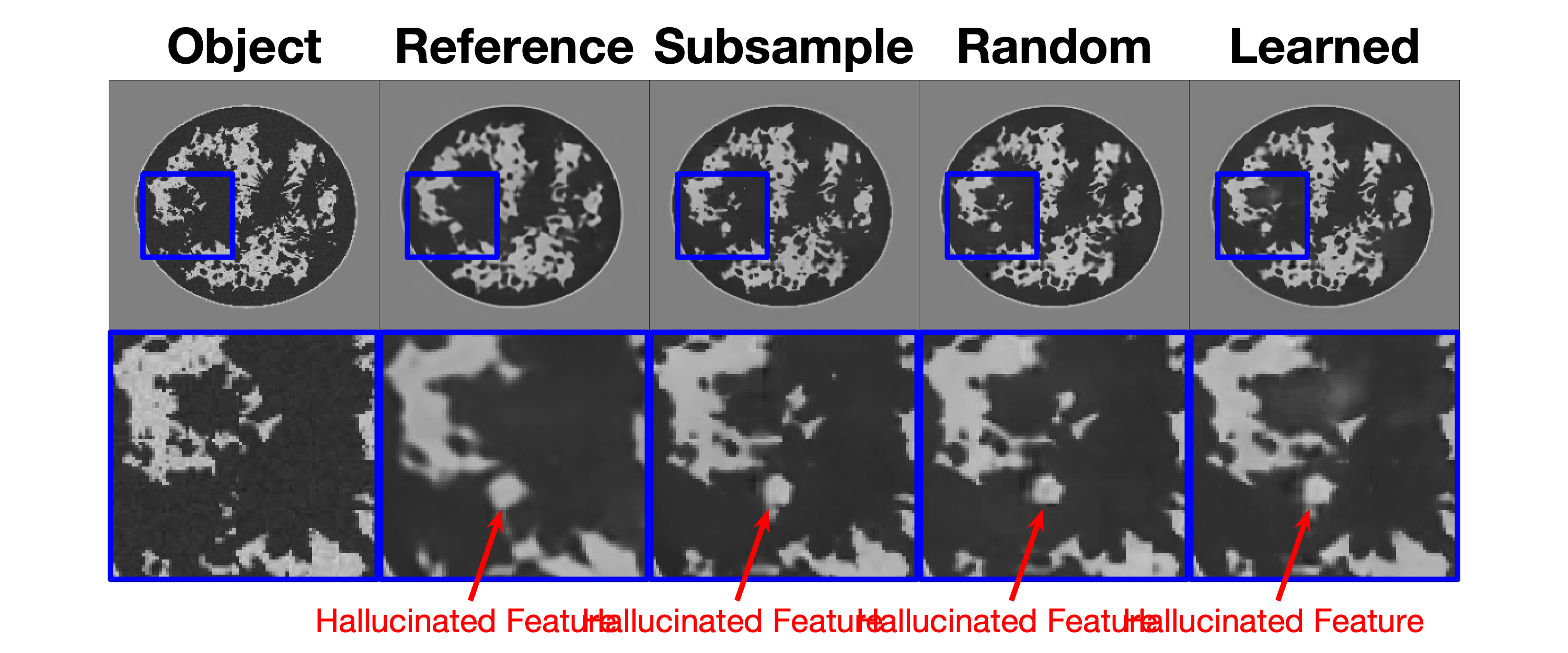}
    \caption{Study 1: Examples of speed of sounds maps reconstructed by each instance of InversionNet in the source encoding study. From left to right are the object and its estimates reconstructed using the reference instance without source encoding, the instance with subsampling, the instance with random source encoding, and that with a learned source encoding. The bottom row is a zoomed-in feature for each image highlighting differences in image resolution and hallucinated features. One notable hallucinated feature is annotated with a red arrow.}
    \label{fig:source_encoding_example}
\end{figure*}

\begin{figure*}[tbh]
    \centering
    \includegraphics[width = 0.325\textwidth]{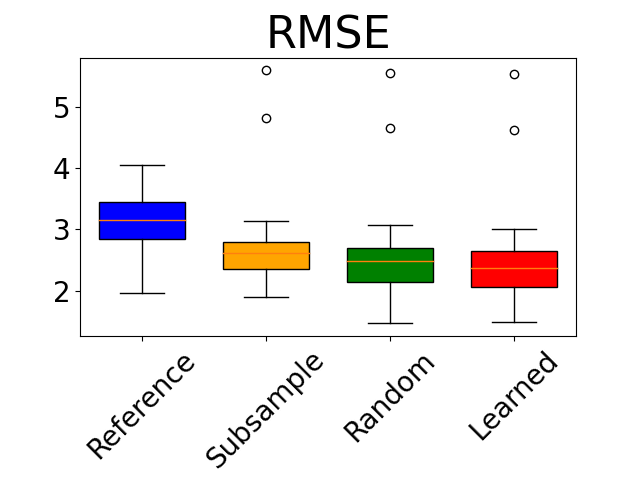}\includegraphics[width = 0.325\textwidth]{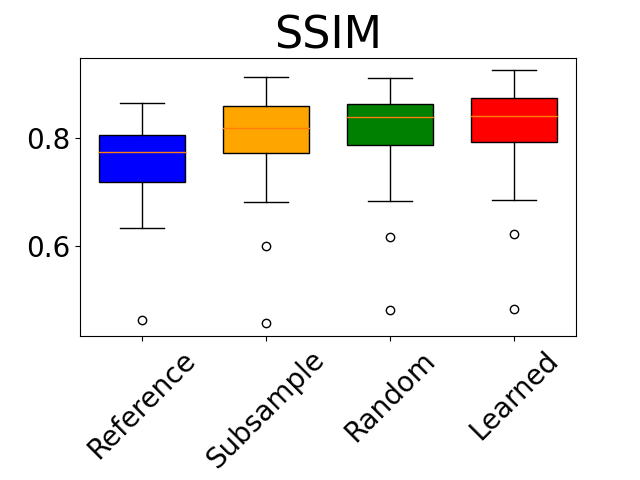}
    \caption{Study 1: Boxplots of RMSEs and SSIMs across the testing set for reconstructions in the source encoding study. The instances of InversionNet utilizing a fixed random source encoder (\emph{Random}) and the learned encoder (\emph{Learned}) demonstrate the lowest ensemble average RMSEs and highest ensemble average SSIMs.}
    \label{fig:se_rmse_boxplots}
\end{figure*}

\subsubsection{Qualitative Assessment} An example reconstruction of a type C NBP from each instance of InversionNet in the source encoding study is shown in Fig. \ref{fig:source_encoding_example}. In this example, the instances of InversionNet utilizing all sources (\emph{Reference}) or subsampling (\emph{Subsample}) have the worst visual appearance (blurred edges, lack of details) while the other instances (\emph{Random}, \emph{Learned}) have better visual appearance (sharper edges, finer details). However, several hallucinated features (i.e. false structures that do not exist in the underlying object but are introduced by the reconstruction method \cite{bhadra21hallucinations}) appear in the images reconstructed by all instances. For example, the estimates  appear to be hallucinating, or overestimating, the presence of a high speed of sound region in the middle of the zoomed region. 
\subsubsection{Quantitative Assessment}  Ensemble average RMSE and SSIM figures of merits computed over the testing set are illustrated in Fig. \ref{fig:se_rmse_boxplots} for each of the four trained instances of InversionNet. These results indicate that the \emph{Random} instance is able to achieve a statistically significant lower ensemble average RMSE than the \emph{Reference} (p-value $3.03e-03\%$) and comparable accuracy to the \emph{Subsample} and \emph{Learned} instances (p-values 23.7\% and 76.5\% respectively). Similarly, the \emph{Random} instance is able to achieve a statistically significant higher ensemble average SSIM than the \emph{Reference} (p-value $0.214\%$) and comparable SSIM to the \emph{Subsample} and \emph{Learned} instances (p-values 45.5\% and 72.4\% respectively). 

\subsubsection{Discussion}
While the \emph{Reference} network has a larger number of trainable parameters than the \emph{Random} network and thus a larger representation power, it performs worse than the \emph{Random} network. Similarly, the \emph{Learned} network has a larger representation power, but yields comparable performance to the the \emph{Random} network. This is possibly due to the fact that all networks were trained on the same number of training examples and for the same number of epochs. It is possible that was a larger training set available and given longer training times, the \emph{Reference} and \emph{Learned} networks would eventually outperform the \emph{Random} network.

%\begin{table*} \centering \caption{Means and Standard deviation for each reconstruction in the source encoding experiment} \begin{tabular}{c|cccccccc} & Reference & Subsample & Random & Trained  \\ \hline  RMSE mean & 3.23 & 3.21 & 2.24 & 2.81\\ RMSE standard deviation & 0.461 & 0.813 & 0.731 & 0.659 \\ \hline SSIM mean & 0.754  & 0.743 & 0.832 & 0.797 \\  SSIM standard deviation & 0.0768 & 0.0797   & 0.0845 & 0.0873\\  \end{tabular} \label{tab:se_stats}\end{table*}

\subsection{Study 2: Task-informed Loss}\label{sec:task}
\begin{figure*}[tbh]
    \includegraphics[width = \textwidth, trim = {0cm 2cm 0cm 0cm},clip]{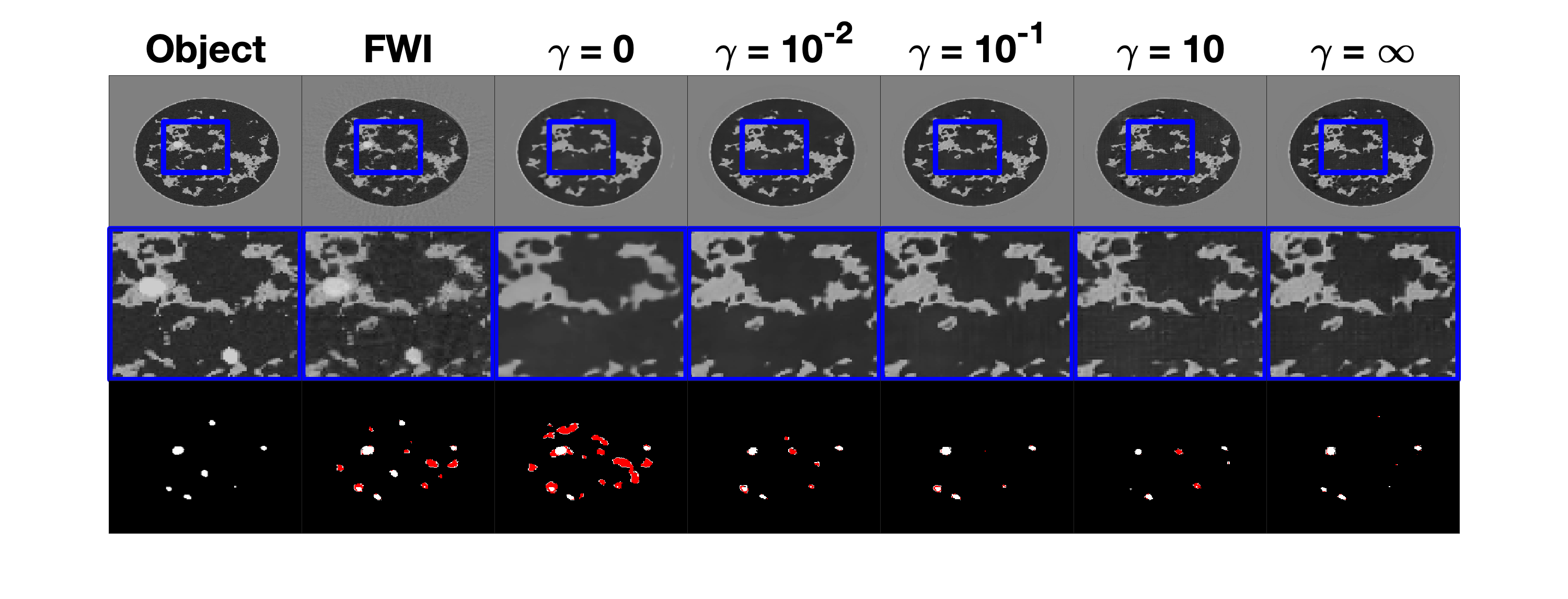}
    \caption{Study 2: Examples of speed of sound maps reconstructed by FWI and various instances of InversionNet trained with increasing weight $\gamma$ of the task informed loss. From left to right is the object, FWI reconstruction, InversionNet reconstructions with $\gamma = 0,10^{-2},10^{-1}, 10$ and $\infty$.  The middle row is a zoomed-in feature for each image highlighting differences in image resolution and detected features. The bottom row is the resulting tumor segmentation with the true tumor material shown in white and the hallucinated tumor materials shown in red. Speed of sound estimates reconstructed using FWI and the instance of InversionNet without task-informed loss shows a large number of hallucinated tumors, while instances of InversionNet trained with $\gamma \geq 10^{-1}$ lead to accurate tumor segmentation masks.}
    \label{fig:task_example}
\end{figure*}

\begin{figure*}[tbh]
    \centering
    \includegraphics[width = 0.325\textwidth, trim = {0cm 0cm 1cm 0cm},clip]{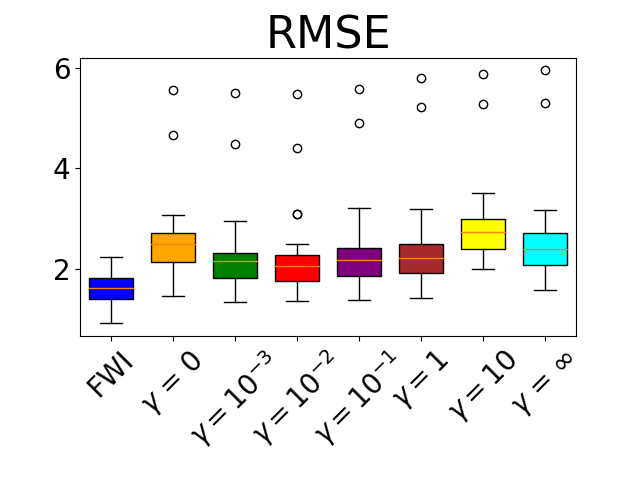} \includegraphics[width = 0.325\textwidth, trim = {0cm 0cm 1cm 0cm},clip]{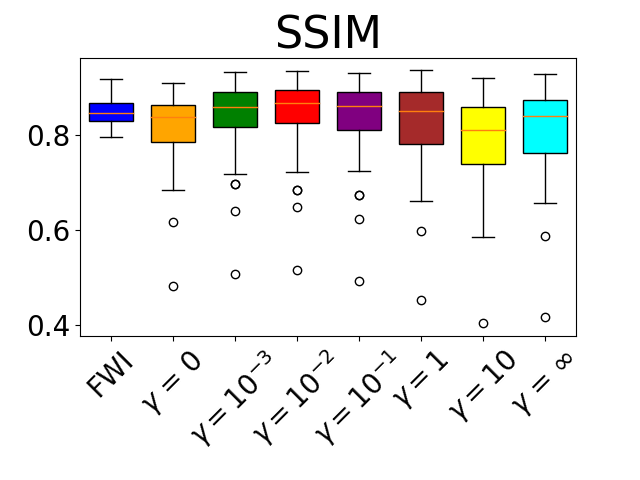} \includegraphics[width = 0.325\textwidth, trim = {0cm 0cm 1cm 0cm},clip]{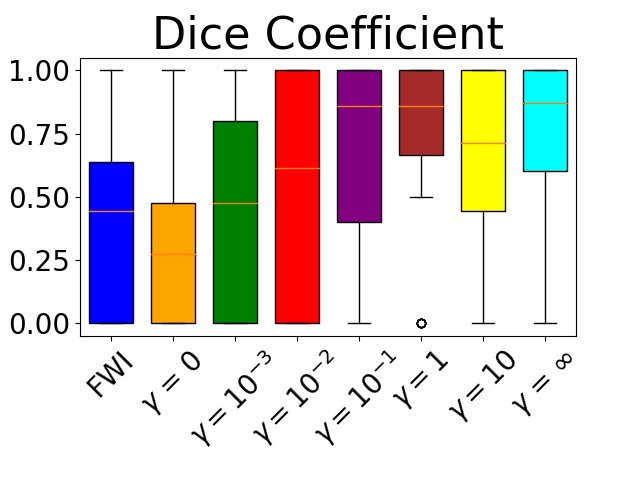}
    \caption{Study 2: Boxplots of RMSEs, SSIMs, and Dice coefficients across the testing set for reconstructions from the task-informed study. Using the task-informed loss reduces RMSEs and increases the SSIMs and Dice coefficient. Best performance is achieved for the task informed weight $\gamma = 10^{-1}$. The iterative FWI method outperforms the InversionNet in terms of RMSE but underperforms in terms of SSIM. With the proper task-informed weight ($\gamma \geq 10^{-1}$) the InversionNet demonstrates better task performance than the iterative FWI methods as quantified by the Dice coefficient.}
    \label{fig:task_rmse_boxplots}
\end{figure*}
\begin{figure}[tbh]
    \centering
    \includegraphics[width =\columnwidth]{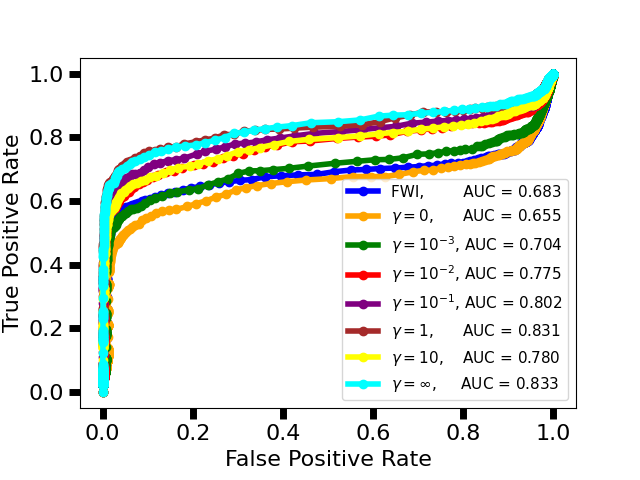}
    \caption{Study 2: Receiver operator characteristic (ROC) curve for FWI and various instances of the task-informed InversionNet. Higher values of the task-informed weight $\gamma$ lead to higher AUC. Instances of InversionNet trained with $\gamma \geq 10^{-3}$ outperform FWI in terms of AUC.}
    \label{fig:roc_curve}
\end{figure}

\subsubsection{Qualitative Assessment} An example reconstruction of a type B NBP from the instances of InversionNet corresponding to $\gamma = 0,10^{-2},10^{-1},1, 10,\infty$ and the FWI reconstruction in the task informed experiment is shown in Fig. \ref{fig:task_example}. The top row of this figure displays the reconstructed images with a window over a selected region of interest, the second row displays a zoomed-in image on the windowed region of interest, and the third row displays the resulting tumor segmentation based on the estimated speed of sound.  Tumor segmentation was obtained by thresholding the output of the U-net observer with a fixed threshold of $0.02$, which was chosen based on a receiver operator curve (ROC) curve analysis. %as described in Appendix \ref{sec:u-net-training}.
The tumor structures are shown in white for tumors that were correctly detected and localized and in red for hallucinated structures.   In this example, there is a notable improvement in image quality as the task-informed weight $\gamma$ increases and peaks at $\gamma = 10^{-1}$. For values of  $\gamma$ equal to or larger than 1, there is a stabilization in visual appearance. The FWI reconstruction exhibits high-frequency artifacts, which can be clearly seen in the zoomed portion. Tumor detection and localization improve as the task weight increases. 

\subsubsection{Quantitative Assessment}   The ROC curves and their respective areas under the curve (AUC) are shown in Fig. \ref{fig:roc_curve}. These ROC curves were used to select the threshold for tumor detection/localization. The selected threshold was 0.02, which corresponds to the upper left corner of the ROC curve. For $\gamma \geq 10^{-3}$, the learned reconstructions demonstrated a higher AUC than the FWI method, with AUC increasing alongside the task weight, with the exception of $\gamma =10$.  

The RMSEs, SSIMs, and Dice coefficients for each of these reconstruction methods across the testing set are illustrated in Fig. \ref{fig:task_rmse_boxplots}. The learned reconstruction methods have a statistically significant higher RMSE compared to the FWI method for all weights $\gamma$ (p-values less that $0.001\%$).
%(5.16e-09\%, 3.80e-04\%, 7.64e-05\%, 3.03e-05\%, 3.15e-13\%, 2.41e-07). 

For certain task weights, there is no statistical difference between the learned reconstruction methods and the FWI method in terms of SSIM (p-values 32.6\%, 57.3\%, 25.1\%, 12.0\% for $\gamma = 10^{-3},10^{-2},10^{-1},1$) while the learned methods underperform with either no or a heavy weight on the task information (p-values 0.601\%, $3.70e-02\%$, 1.68\% for $\gamma=0,10,\infty$).  For low task weights($\gamma \leq 10^{-2}$), the learned reconstruction methods perform comparably to the FWI method in terms of Dice coefficient (p-values $geq 11\%$)
%(p-values 28.9\%, 51.9\$ 11.1\% for $\gamma = 0,10,10^{-3},10^{-2}$) 
 and statistically significantly better for higher task weights 
(p-values $\leq 0.215\%$ for $\gamma \geq 10^{-1}$). 
 %(p-values 0.215\%, 1.72e-02\%,2.23e-02\%,  for $\gamma = 10^{-1},1,10,\infty$).

\subsubsection{Discussion} 

Continuation over $\gamma$ was needed in this study to ensure that InversionNet converged to a ``good'' local minima using the task-informed loss function. Preliminary experiments showed that training a learned reconstruction with a large task weight and without a good initialization led to very poor image quality and task performance, most likely due to the nonlinearity introduced by the numerical observer. A possible limitation of this study is that no regularization was used in the FWI reconstruction. However, the use of densely samples data, absence of modeling error (discretization inverse crime), and high signal to noise ratio mitigate this limitation.

%\begin{table*} \centering \caption{Means and Standard deviation for each reconstruction in the task-informed experiment} \begin{tabular}{c|c|ccccccc} & FWI & $\gamma = 0$ & $\gamma = 10^{-3}$ & $\gamma = 10^{-2}$ & $\gamma = 10^{-1}$ & $\gamma = 1$ & $\gamma = 10$ & $\gamma = \infty$ \\ \hline  RMSE mean & 1.43 & 2.24 & 2.09 & 2.10 & 2.12 & 2.43 &  2.45 & 2.38 \\  RMSE standard deviation & 0.828 & 0.734 & 0.729 & 0.714 & 0.774 & 0.789 & 0.799 & 0.808 \\  \hline  SSIM mean & 0.886 & 0.831 & 0.844 & 0.845 & 0.844 & 0.820 & 0.817 & 0.821 \\ SSIM standard deviation & 0.0571 & 0.0845 & 0.0829 & 0.0837 & 0.0886 & 0.0921 & 0.998 & 0.960 \\ \hline Dice mean & 0.550 & 0.322 & 0.445 & 0.587 & 0.720 & 0.695 & 0.717 & 0.705 \\ Dice standard deviation & 0.407 & 0.351 & 0.395 & 0.392 & 0.357 & 0.374 & 0.377 & 0.374 \end{tabular}\label{tab:task_stats} \end{table*}

\subsection{Study 3: Robustness w.r.t. Noise}
\label{sec:robust}
\begin{figure*}[tbh]
    \includegraphics[width = \textwidth, trim = {0cm 2cm 0cm 0cm},clip]{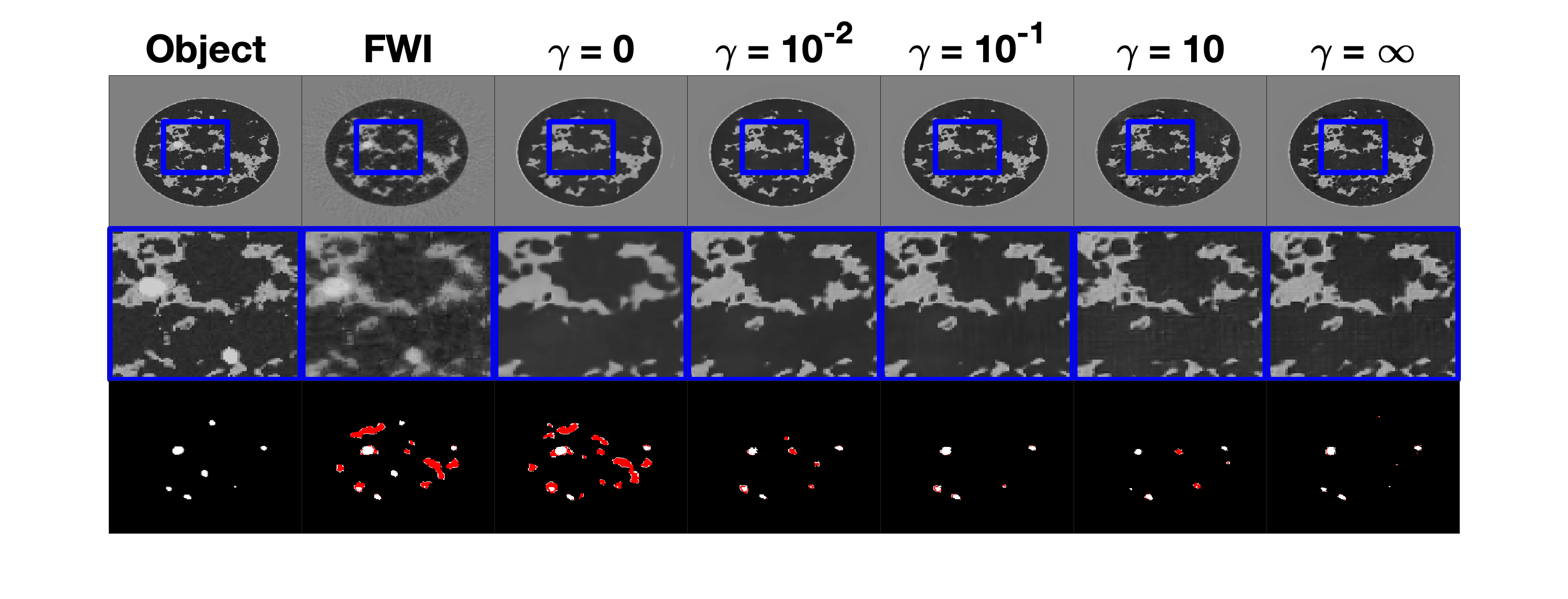}
    \caption{Study 3: Examples of speed of sound maps reconstructed using FWI and various instances of InversionNet in the robustness study. From left to right is the object, FWI reconstruction, InversionNet reconstructions with task weight parameter $\gamma = 0,10^{-3},10^{-2},10^{-1}, 1, 10$ and $\infty$.  The middle row is a zoomed-in feature for each image highlighting differences in image resolution and detected features. There is no visually clear degradation in the quality in reconstructions despite the increased level of measurement noise. The bottom row is the resulting tumor segmentation with the true tumor material shown in white and the hallucinated tumor materials shown in red. Comparing the visual appearance of reconstructed images and segmentation masks with those shown in Fig. \ref{fig:task_example} for the same object demonstrates the robustness of InversionNet with respect to measurement noise.}
    \label{fig:rob_example}
\end{figure*}

\begin{figure*}[tbh]
    \centering
    \includegraphics[width = 0.325\textwidth, trim = {0cm 0cm 1cm 0cm},clip]{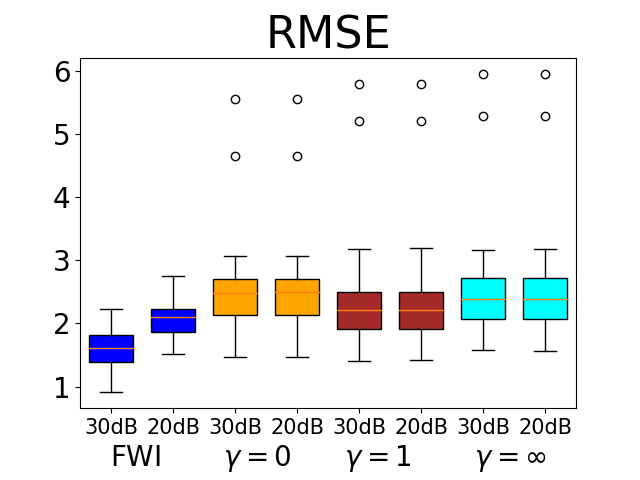}\includegraphics[width = 0.325
    \textwidth, trim = {0cm 0cm 1cm 0cm},clip]{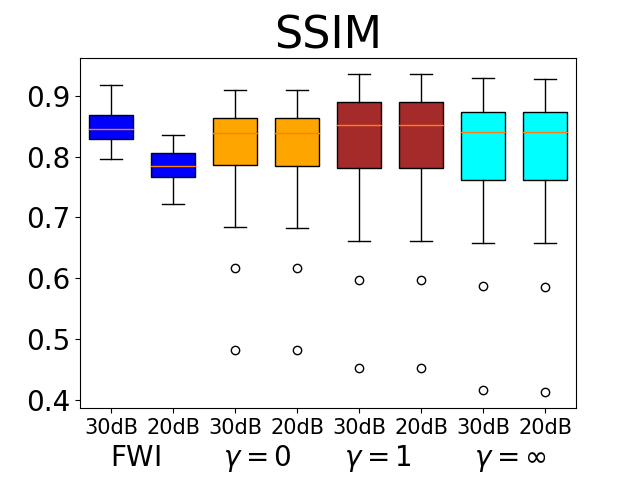}\includegraphics[width = 0.325\textwidth, trim = {0cm 0cm 1cm 0cm},clip]{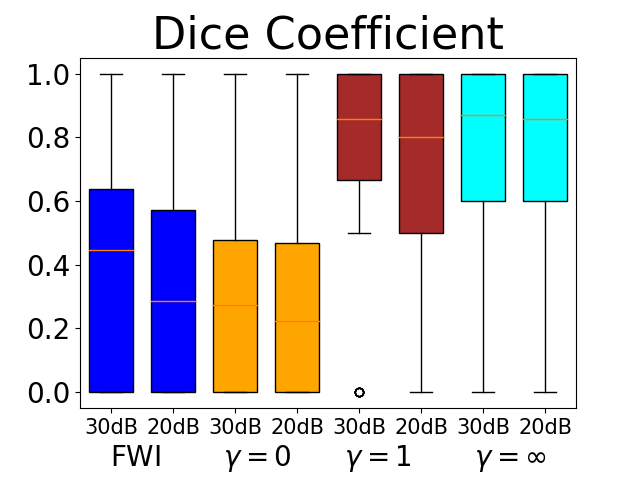}
    \caption{Study 3: Boxplots of RMSEs, SSIMs, and Dice coefficients across the testing set for reconstructions from robustness study. Each color corresponds to an instance of InversionNet trained with a specified task weighting $\gamma$ and tested with two different levels of measurement noise. The increased level of noise results in a slight increase in RMSE and decrease in SSIM and Dice coefficient.}
    \label{fig:rob_rmse_boxplots}
\end{figure*}

\begin{figure}
    \centering
    \includegraphics[width =\columnwidth]{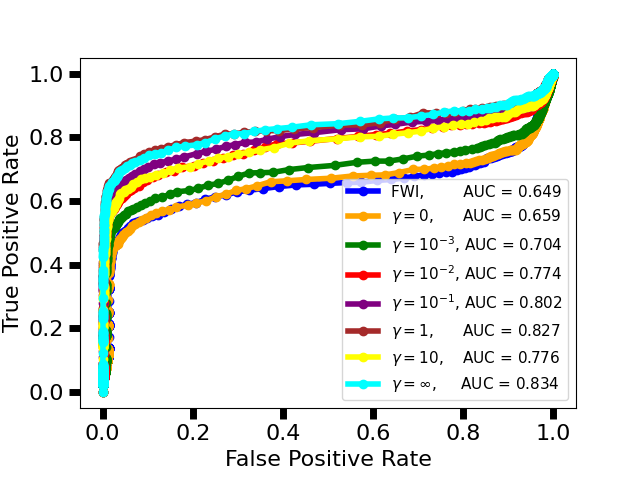}
    \caption{Study 3: Receiver operator characteristic (ROC) curve for task-informed, learned reconstruction methods with increased noise level. All instances of InversionNet achieved a higher AUC than FWI. Compared with Fig. \ref{fig:roc_curve}, AUC for all methods decreases slightly with the higher level of noise.} 
    \label{fig:rob_roc_curve}
\end{figure}

\subsubsection{Assessment} An example reconstruction of a type B NBP, the same from the previous study, from each instance of InversionNet from the robustness experiment is shown in Fig. \ref{fig:rob_example}. In this example there is not a noticeable drop in image quality or a clear difference in tumor segmentation compared to the low-noise reconstructions in the Section \ref{sec:task}.
The ROC curve for each reconstruction is plotted in Fig. \ref{fig:rob_roc_curve}, with the corresponding AUC shown in the legend. The increased level of noise results in a very slight decrease in AUC for all reconstruction methods, thus demonstrating observer robustness with respect to noise. Furthermore, for $\gamma \geq 10^{-1}$ the learned reconstruction methods demonstrate a higher AUC than the FWI method. Figure \ref{fig:rob_rmse_boxplots} displays box plots of RMSEs, SSIMs, and Dice coefficients achieved by the proposed method with $\gamma=0, 1, \infty$ as well as by FWI. For the FWI reconstructions, a statistically significant increase in RMSE and decrease in SSIM was noted in the high noise compared to low noise (Case study 2) cases (p-values: $9.82e-09\%$ and  $9.26e-15$, respectively). No statically significant change in RMSE and SSIM was noted for the learned reconstructions with $\gamma = 0,1,\infty$ (all p-values $\geq 95\%$).
Moreover, for each reconstruction method (FWI, $\gamma=0,1,\infty$), the reduction in  Dice coefficient  between the low noise and high noise reconstructions is statistically insignificant (p-values $44.1\%,84.1\%,75.1,$ and  $96.1\%$).

\subsubsection{Discussion} Performance in terms of RMSE, SSIM, Dice coefficient, and AUC for the learned reconstruction methods decreases very slightly, statistically insignificantly, with the presence of increased noise and maintains its comparative performance with respect to the FWI method. This study shows that the proposed learned reconstruction methods are robust with respect to increased noise.

%begin{table*} \centering \caption{Means and Standard deviation for each reconstruction in the robustness experiment} \begin{tabular}{c|ccccccc} & $\gamma = 0$ & $\gamma = 10^{-3}$ & $\gamma = 10^{-2}$ & $\gamma = 10^{-1}$ & $\gamma = 1$ & $\gamma = 10$ & $\gamma = \infty$ \\ \hline  RMSE mean                & 2.48 & 2.80 & 2.32 &  2.29 & 2.57 & 2.59 & 2.49 \\  RMSE standard deviation  & 0.685 & 0.645 & 0.703 & 0.749 & 0.764 & 0.791 & 0.782 \\ \hline SSIM mean & 0.815 & 0.794 & 0.831 & 0.833 & 0.810 & 0.807 & 0.813 \\ SSIM standard deviation  & 0.0869 & 0.0800 & 0.0849 &  0.0912 & 0.0949 & 0.105 & 0.0971 \\  \hline Dice mean  & 0.274 & 0.350 & 0.609 & 0.664 & 0.597 & 0.705 & 0.718 \\ Dice standard deviation & 0.321 & 0.378 & 0.392 &  0.383 & 0.399 & 0.377 & 0.359 \\ \end{tabular} \label{tab:rob_stats} \end{table*}

\subsection{Study 4: Generalizability w.r.t. underrepresented groups}
\label{sec:gen}
\begin{figure*}[tbh]
    \includegraphics[width = \textwidth, trim = {0cm 2cm 0cm 0cm},clip]{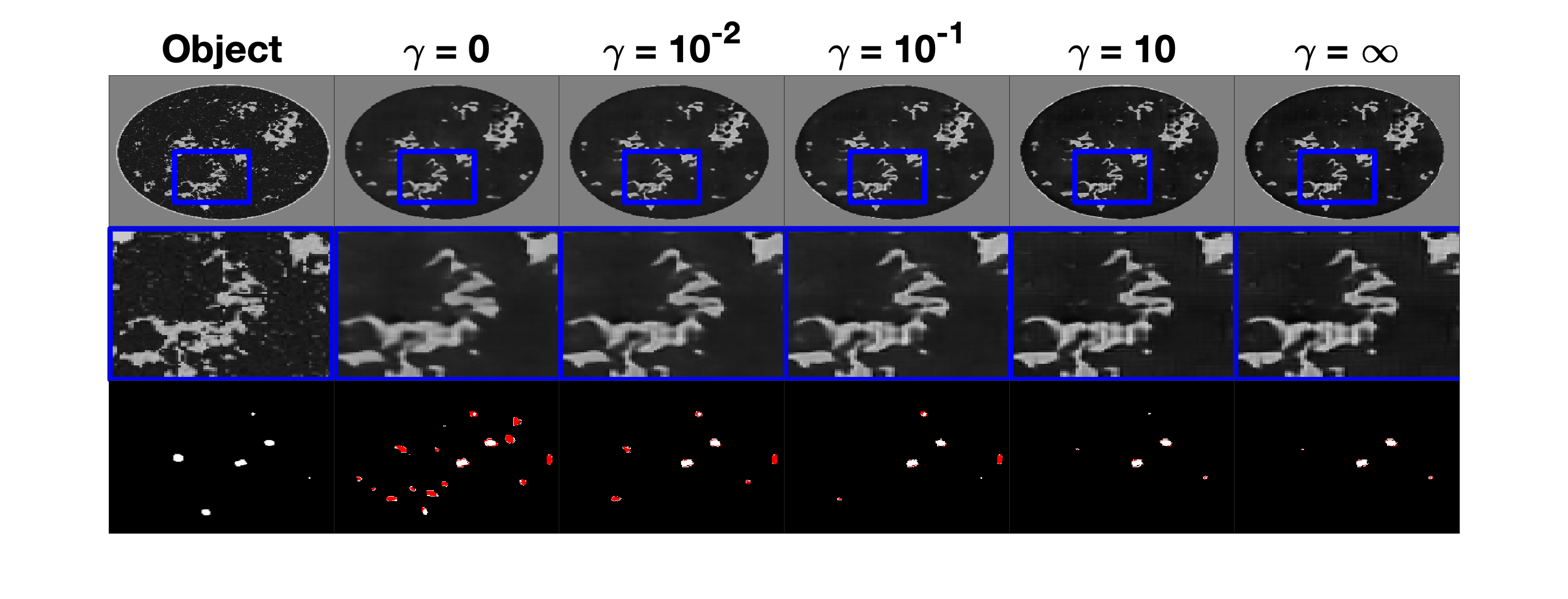}
    \caption{Study 4: Examples of speed of sounds maps reconstructed by each instance InversionNet in the generalization study. From left to right is the object and InversionNet reconstructions with $\gamma = 0,10^{-2},10^{-1}, 10$ and $\infty$.  The middle row is a zoomed-in feature for each image highlighting differences in image resolution and reconstructed features. The bottom row is the resulting tumor segmentation with the true tumor material shown in white and the hallucinated tumor materials shown in red. Image quality and tumor detection noticeably decrease when the testing set consists of an underrepresented population in the training set but does not completely degrade and is still relatively accurate.}
    \label{fig:gen_example}
\end{figure*}

\begin{figure*}[tbh]
    \centering
    \includegraphics[width = 0.325\textwidth, trim = {0cm 0cm 1cm 0cm},clip]{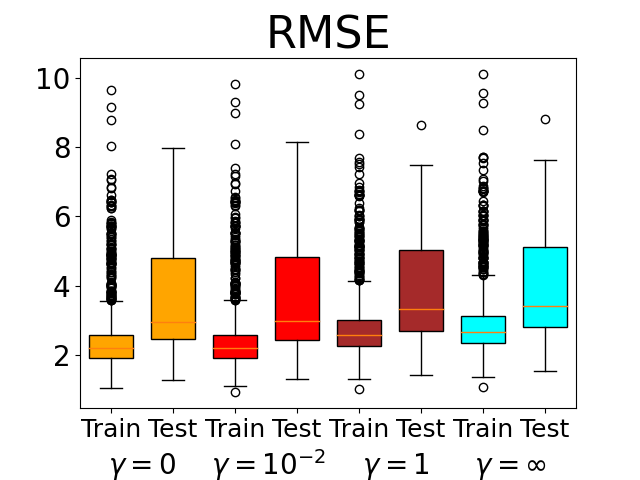}\includegraphics[width = 0.325\textwidth, trim = {0cm 0cm 1cm 0cm},clip]{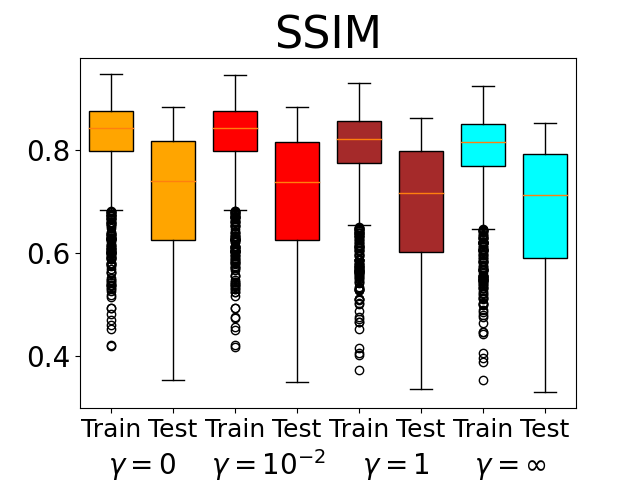}\includegraphics[width = 0.325\textwidth, trim = {0cm 0cm 1cm 0cm},clip]{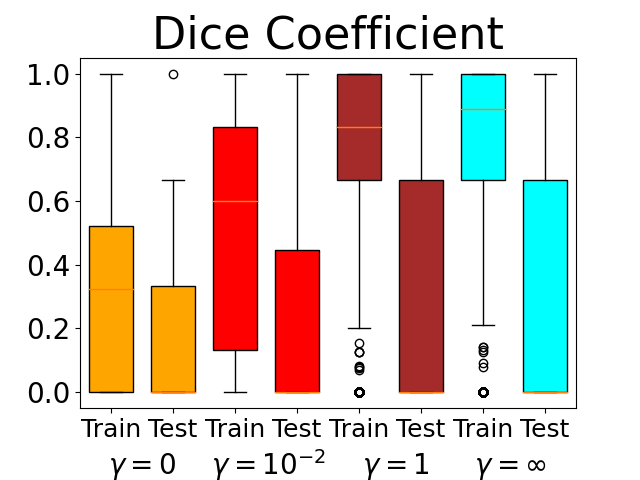}
    \caption{Study 4: Boxplots of RMSEs, SSIMs, and Dice coefficients across the testing set for reconstructions from generalization study. The testing set generated from an underrepresented group demonstrates an increase in RMSE and decrease in SSIM and Dice coefficient.}
    \label{fig:gen_rmse_boxplots}
\end{figure*}

\begin{figure}[tbh]
    \centering
    \includegraphics[width =\columnwidth]{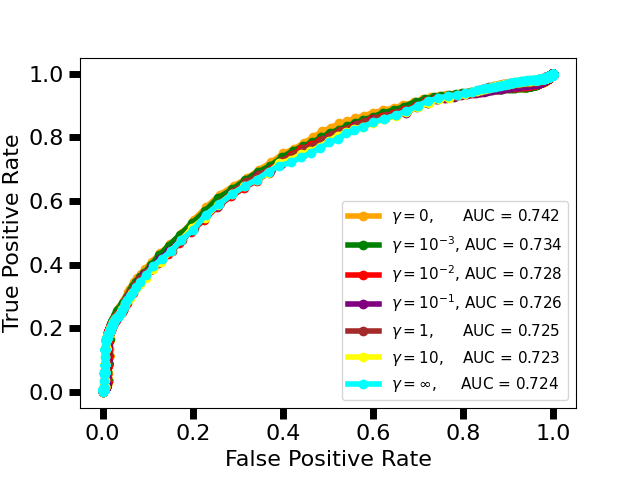}
    \caption{Study 4: Receiver operator characteristic (ROC) curve for task-informed, learned reconstruction methods on underrepresented population. Compared to Fig. \ref{fig:roc_curve}, this study displays a clear decrease in observer performance on an underrepresented population, with a no improvement in AUC as $\gamma$ increases.} 
    \label{fig:gen_roc_curve}
\end{figure}

\subsubsection{Assessment} An example reconstruction from each instance of InversionNet from the generalization experiment is shown in Fig. \ref{fig:gen_example}. In these images there is a drop in image quality compared to the other experiments and the accuracy of the tumor segmentation decreases. The ROC curve for each reconstruction method is plotted in Fig. \ref{fig:gen_roc_curve}, with the corresponding AUC shown in the legend. This displays that the task-informed objective does not improve  task performance for underrepresented reconstructions.  For a select few of these reconstruction methods for both  the training and testing sets, the RMSEs, SSIMs, and Dice coefficients  across the testing set are illustrated in Fig. \ref{fig:gen_rmse_boxplots}. The boxplots display a significant decrease in image quality and task performance for underrepresented data.

\subsubsection{Discussion} 
In addition to the experiments reported here, several instances of InversionNet were trained on a data set consisting of only types B, C, D NBPs and then evaluated on an out-of-distribution testing set consisting only of type A NBPs. Results showed poor generalization performances in this case and are not reported here. This decrease in performance was expected as it is known that deep learning methods that map between different input/output domains tend to perform worse, be more difficult to train, and generalize poorly compared to methods that map between input/outputs in the same domain \cite{pang2021image}. This means that InversionNet is at a disadvantage for achieving accurate performance compared to other methods that combine a learned image (or data) translation/correction and an approximated physics-based mapping between the data and image domains \cite{jeong2023deep, feng2022intriguing}.

%\begin{table*} \centering \caption{Means and Standard deviation for each reconstruction in the generalization experiment} \begin{tabular}{c|ccccccc} & $\gamma = 0$ & $\gamma = 10^{-3}$ & $\gamma = 10^{-2}$ & $\gamma = 10^{-1}$ & $\gamma = 1$ & $\gamma = 10$ & $\gamma = \infty$ \\ \hline  RMSE training mean                & 1.64 & 1.64 & 1.66 & 1.78 & 1.91 & 1.96 & 1.98 \\  RMSE training standard deviation  & 0.656 & 0.655 & 0.657 & 0.668 & 0.712 & 0.730 & 0.735\\ RMSE testing mean                & 3.22 & 3.23 & 3.23 & 3.35 & 3.45 & 3.47 & 3.49 \\RMSE testing standard deviation  & 1.68 & 1.70 & 1.69 & 1.73 & 1.74  & 1.73 & 1.73 \\ \hline  SSIM training mean                & 0.883 & 0.884 & 0.883 & 0.876 & 0.868 & 0.864 & 0.862 \\ SSIM training standard deviation  & 0.0619 & 0.0618 & 0.0616 & 0626 & 0.0664 & 0.0676 & 0.0682 \\ SSIM testing mean           0.742 & 0.742 & 0.742 &   0.734 & 0.728 & 0.724 & 0.723 \\  SSIM testing standard deviation  & 0.132 & 0.133 & 0.132 & 0.135 & 0.134 & 0.133 & 0.133 \\ \hline Dice training mean               & 0.577 & 0.638 & 0.752 & 0.834 & 0.869 & 0.875  & 0.884 \\ Dice training standard deviation & 0.322 & 0.317 & 0.293 & 0.251  & 0.231 & 0.221 & 0.207 \\ Dice testing mean               & 0.271 & 0.288 & 0.339 & 0.361 & 0.389 & 0.383 & 0.397 \\ Dice testing standard deviation & 0.302 & 0.322 & 0.360  & 0.397 & 0.406 & 0.411 & 0.415 \\\end{tabular} \label{tab:gen_stats}\end{table*}

\section{Conclusion}\label{sec:conclusion}

This work presents a deep-learning image reconstruction method for ultrasound computed tomography (USCT). 
In particular, the InversionNet architecture, originally proposed for seismic imaging, was extended to produce quantitatively accurate speed of sound maps of breast tissues from simulated USCT data, without the computational burden of model-based iterative methods, such as full waveform inversion (FWI).
The proposed deep learning image reconstruction method was illustrated in four virtual imaging studies using a large set of anatomically and physiologically realistic numerical breast phantoms. 

The first study assessed different source encoding approaches for InversionNet. The source encoding study showed that random source encoding results in lower RMSE and higher SSIM compared to a subsampling. Furthermore, the network trained using random source encoding performed  better than the reference network (no source encoding) and the one with learned source encoder. A reason for this is that all networks were trained on the same training set and for the same number of epochs. While it is expected that---given a sufficiently large training set and training time---using no source encoding or a trained encoder would eventually lead to better performance, this study demonstrates that a fixed random source encoder is able to reduce data complexity and simplify training, leading to improved accuracy when training data is limited. 

The second study assessed the role of a task-informed loss function in training InversionNet and its effect on task performance. In this study, increasing the weight of the task-informed loss in the loss function used during training increased image quality both in terms of RMSE and SSIM until it was competitive with an FWI reconstruction despite being three orders of magnitude faster than FWI. Furthermore, network trained with a task-informed objective demonstrated better task performance in terms of Dice coefficient compared to the FWI reconstruction. Broadly, this study shows that a learned reconstruction can be tailored for a specific task and that task information can improve image quality. 

The third study assessed the proposed method
 robustness with respect to additive noise levels. In this study, the learned FWI networks trained in the second study were used to reconstruct unseen data which were corrupted using a ten times higher noise level. This study exhibited minimally diminished accuracy compared to the second study. These results demonstrate that the proposed method has robustness with respect to measurement noise. 

The fourth study assessed InversionNet's ability to reconstruct out-of-distribution images. In this study, an instance of InversionNet was trained using a dataset for which BI-RADS type A (fatty breast) NBPs were severely underrepresented, with only one-fifth the number of examples as the other categories. The network achieved only slightly diminished accuracy (compared with that achieved in the second study) and then tested on a testing set consisting of only BI-RADS type A NBPs. These results demonstrate that InversionNet is able to generalize for reconstruction on underrepresented populations in the training sets. Nevertheless, the diminished accuracy of this underrepresented population highlights the need for a large distribution of representative training images for highly accurate results. 

In summary, this work established the feasibility of employing a learned FWI reconstruction method employing CNNs from USCT data and demonstrated reduced computational burden and the ability to leverage task-specific information.  Future work will include the application of these learned reconstruction methods to clinical data. For this application, several additional challenges will need to be addressed. Primarily, the forward model implemented will be extended to a 3D wave physics model and a 3D ring array USCT imaging system  \cite{Li23}. Additionally, future works will address the lack, rarity, or ground truth images by implementing a self-supervised training method \cite{jin2022unsupervised}.

\section*{Acknowledgments}
 This work was co-funded by the Los Alamos National Laboratory~(LANL) - the Center for Space and Earth Science  and Laboratory Directed Research and Development program under project number 20200061DR. Luke Lozenski, Fu Li, Mark Anastasio, and Umberto Villa would like to acknowledge support from NIH under award number R01 EB028652.

\appendices
\section{Training a U-Net Observer}
\label{sec:u-net-training}

The numerical observer used in the tumor detection and localization task study utilized a U-Net architecture. Mathematically, the U-Net observer can be represented as an operator $\boldsymbol{P} = \Psi_\xi(\boldsymbol{C})$, with trainable parameters $\wts \in R^{p'}$ that maps a speed of sound image $\boldsymbol{C} \in \mathbb{R}^q$ to an image $\boldsymbol{P} \in [0,1]^Q$ those pixel values represent the probability that a tumor is present in that location. %The lesions, of random but known shape, are inserted at plausible and known locations when constructing the NBPs. Therefore the indicator function of lesion regions is available for training.
The U-Net observer was then trained in a supervised manner with a pixelwise cross-entropy loss function that compares the output of $\Psi_\xi$ with the truth tumor segmentation maps. A training set consisting of all the 1,435 available speed of sound NBPs and corresponding tumor segmentation maps was used to train the U-net observer. %This observer demonstrated very accurate tumor classification across the set of numerical phantoms.
The area under the receiver operator characteristic curve was 0.941. %with the tumor-wise detection rate on the x-axis and pixel-wise false positive rate on the y-axis for every phantom in the training and testing sets, %is displayed in Fig. \ref{fig:obs_curve}. The area under the curve (AUC) was 0.941. It is important to note that a rigorous validation of the observer itself is outside the scope of this work. Instead, this work assumes that a validated numerical observer for the specific task is already given and focuses on training learned reconstruction methods able to preserve features of the image that are relevant to the observer. 

%\begin{figure}
%    \centering
%    \includegraphics[width =\columnwidth]{Figures/obs_roc_curve.png}
%    \caption{Receiver operator characteristic (ROC) curve for task-informed, U-Net tumor observer.}
%    \label{fig:obs_curve}
%\end{figure}

\bibliographystyle{IEEEtran}
\bibliography{local, references}

\end{document}